# Max Bense en visionnaire : de l'entropie à la dialectique des images programmées

*Max Bense as a Visionary: from Entropy to the Dialectics of Programmed Images*

**Résumé**

En 1960 à Stuttgart, le philosophe Max Bense publie l'ouvrage *Programmer le beau* [*Programmierung des Schönen*]. Bense cherche dans la cybernétique des concepts scientifiques et inaugure la pensée de la programmation dans le domaine de la littérature. Son esthétique de l'information marque toute une génération de scientifiques et d'artistes – dont le Cercle de Stuttgart qui s'empare de *la nouvelle esthétique* pour faire émerger les premières images artistiques programmées. Max Bense est-il un visionnaire ? De quelle façon révolutionne-t-il le monde des images ? L'article discute de la cybernétique qui inspire Bense : une science des probabilités en rupture avec les principes de la physique newtonienne. Par ailleurs, dans les années soixante, Max Bense lance avec Elisabeth Walther la revue expérimentale *Rot* qui consacre ses pages à la poésie concrète et aux premières images générées par ordinateur de Georg Nees. Comme le défend Frieder Nake à travers son œuvre pionnière et sa théorie, ces images opposent le visible et le calculable. Cette dialectique ouvre à une réflexion critique sur l'image algorithmique en art et en science.

Mots clefs : esthétique, information, cybernétique, entropie, programmation, texte, image, algorithme, Computer Art, synthèse, visualisation

**Résumé en anglais**

In 1960 in Stuttgart, Max Bense published the book *Programming the Beautiful* [*Programmierung des Schönen*]. Bense looks in cybernetics for scientific concepts and instigates the thought of programming in the field of literature. His information aesthetics influences a whole generation of scientists and artists - including the Stuttgart Circle, which takes hold of *the new aesthetics* to carry out the first programmed artistic images. Is Max Bense a visionary? How is he revolutionizing the world of images? The article discusses the cybernetics that inspired Bense: a science of probability that contrasts with the principles of Newtonian physics. Moreover, in the sixties, Max Bense, together with Elisabeth Walther, launched the experimental magazine *Rot*, which devoted its pages to the concrete poetry and the first computer-generated images of Georg Nees. As Frieder Nake defends through his pioneering work and theory, these images oppose the visible and the computable. This dialectic opens to a critical thinking on the algorithmic image in art and science.

Keywords: aesthetics, information, cybernetics, entropy, programming, text, image, algorithm, Computer Art, synthesis, visualization

**Plan**

1. Théorie de l'information : un pont entre physique et esthétique
2. Programmer ? Du langage scientifique au langage littéraire
3. Nouvelles images, pour une esthétique générative
4. Frieder Nake : une dialectique entre algorithme et image
5. Réflexions sur l'image en art et en science

**Auteur**

Gaëtan Robillard est artiste visuel et professeur associé à l'Université Gustave Eiffel. Il est aussi doctorant du laboratoire Arts des images et art contemporain de l'Université Paris 8, et artiste-chercheur associé au Laboratoire des intuitions de l'ESAD TALM-Tours. Ses recherches théoriques portent sur l'émergence de l'esthétique générative en Europe dans les années soixante et sur les effets de la pensée algorithmique sur l'image numérique au XXI$^e$ siècle. Il développe également une pratique de l'installation et des images dans lesquelles entrent en jeu l'histoire, les mathématiques et les sciences de l'environnement. En 2021, il est sélectionné en résidence par le nouveau consortium MediaFutures de l'Union Européenne, dans le cadre du programme de recherche et d'innovation Horizon 2020. Son travail est régulièrement présenté en France et à l'international, dans des instituts tels que la fondation art-science Le Laboratoire (Cambridge, US), le Centre Pompidou de Metz, la Biennale d'art de Lyon, le Festival international de cinéma FID (Marseille), l'Institut Konrad Lorenz (Vienne), le Palais de Tokyo (Paris), Akbank Sanat (Istanbul), CCS Bard Hessel Museum (New-York) et le Pearl Art Museum (Shanghaï).

**Max Bense en visionnaire : de l'entropie à la dialectique des images programmées**

Max Bense (1910-1990) a étudié les mathématiques, la physique, la géologie et la philosophie à l'université de Bonn où il a obtenu son doctorat en philosophie et en science en décembre 1937[1]. Philosophe, essayiste, poète, éditeur, collectionneur, il est aussi devenu une figure médiatique s'opposant à la société conservatrice dans l'Allemagne d'après-guerre. En 1960, alors qu'il est maître de conférence à l'université de Stuttgart, Max Bense publie *Aesthetica IV : Programmer le beau* [*Programmierung des Schönen*][2]. Ce quatrième livre achève un développement philosophique que Bense démarre en 1954, poursuivant la voie d'une esthétique rationnelle. Le premier livre *Aetshetica I* entend constituer une « conscience cartésienne dans le domaine de l'art », *Aesthetica II* introduit la théorie esthétique de l'information et *Aesthetica III* traite de l'esthétique et la civilisation, comprenant sa réalité technique. *Aesthetica IV* se consacre à la possibilité d'une littérature générée par des programmes et dont les méthodes seraient dérivées des sciences physiques. L'auteur y défend le concept de *textes visuels* dont l'esthétique se développe non seulement comme « flux sémiotique unidimensionnel », mais aussi comme flux qui peut être *vu* tel un évènement sur une surface. En 1965, les quatre livres ont été réédités et rassemblés dans un seul livre – l'*Introduction à la nouvelle esthétique*[3] – accompagnée pour sa version originale (outre les diagrammes) de quatre images mélangeant iconographie scientifique et artistique.

Allant de la sémiotique à la théorie de l'information, et du texte à l'image, la théorie esthétique de Max Bense influence largement la scène pionnière du Computer Art des années soixante. Ses publications peuvent être suivies dans les principaux évènements artistiques internationaux qui font écho à l'arrivée de l'ordinateur dans l'art[4]. Présenté aux côtés de Umberto Eco ou Abraham Moles, le philosophe Allemand s'institue dans les publications internationales comme l'un des précurseurs de ce courant. Pourtant, même si Max Bense contribue au monde des arts visuels et inaugure l'esthétique générative – une première théorie des images algorithmiques, son ouvrage le plus marqué par une réflexion sur le sujet de la programmation informatique, le livre IV de *Aesthetica*, est un ouvrage essentiellement consacré à la littérature et à la théorie du texte.

---

1 Elisabeth WALTHER, « Max Bense's Informational and Semiotical Aesthetics », *Stuttgarter Schule*, 2000. [consulté le 2 novembre 2019]. https://www.stuttgarter-schule.de/bense.html
2 Max BENSE, *Aesthetica IV : Programmierung des Schönen. Allgemeine Texttheorie und Textästhetik*, Baden-Baden ; Krefeld, Agis, 1960.
3 Max BENSE, *Aesthetica : Introduction à la nouvelle esthétique*, Paris, Les éditions du cerf, 2007 [éd. orig. *Aesthetica : Einführung in die neue Aesthetik*, Baden-Baden, Agis, 1965].
4 Margit ROSEN (dir.), *A Little-Known Story about a Movement, a Magazine, and the Computer's Arrival in Art: New Tendencies and Bit International, 1961-1973*, Cambridge, MA: Mit Press, ZKM Karlsruhe, 2011.

Aujourd'hui, les algorithmes sont omniprésents dans la sphère des communications, et plus particulièrement dans la façon dont l'image artistique ou scientifique nous apparait à travers des programmes et des calculs (interfaces, jeux vidéo, simulations scientifique, réalités virtuelles, etc.). Max Bense était-il un visionnaire ? A-t-il anticipé *La révolution algorithmique*[5] qui a eu eu lieu non seulement dans la société de l'information mais aussi dans le monde des images ? Si sa théorie inspire à des mathématiciens et artistes pionniers du Computer Art, comme Georg Nees et Frieder Nake, la réalisation d'œuvres visuelles grâce à la programmation informatique, cette théorie est-elle encore opérante pour les images scientifiques ou artistiques du début du XXIe siècle ?

Pour tenter de répondre à ces questions, nous analyserons quelques-unes des théories esthétiques de Max Bense ainsi que des images qu'il a fait paraître ou qui succèdent à ses théories. Ces théories, ces images et leurs portées, sont examinées selon quatre axes : théorie scientifique, programmation, esthétique générative (dans l'édition *Rot*), et dialectique de l'image programmée chez Frieder Nake. La cinquième et dernière partie ouvre à une réflexion critique sur l'image en art et en science, notamment par l'examen de visualisations recueillies sur le terrain des sciences du climat.

**Théorie de l'information : un pont entre physique et esthétique**

À la fin des années cinquante, Max Bense cultive une vision, celle d'une civilisation technique[6] à venir. Pour se réaliser, celle-ci met à profit « certains concepts, tels qu'ils ont été élaborés par Wiener, Shannon, Carnap, et d'autres »[7]. Ce qui attire Bense dans le courant cybernétique du mathématicien Norbert Wiener et dans la théorie de l'information (Shannon, 1948), c'est la possibilité de fonder une esthétique des « temps modernes », afin d'« aborder […] rationnellement et conformément à l'époque qui est la nôtre, les questions et les résultats relatifs au sujet [esthétique] ».

---

5 Voici comment en 2004 le Centre d'art et de technologie des médias de Karlsruhe pouvait titrer l'une de ses grandes expositions temporaire. À travers cette présentation d'une Histoire de l'art interactif, les curateurs dont Peter Weibel, font le constat que si une révolution s'annonce en général dans le bruit et la fureur, celle des algorithmes au contraire est passée derrière nous sans que personne ne la remarque tout à fait.
6 Max BENSE, *Technische Existenz: Essays*, Stuttgart, Deutsche Verlags-Anstalt, 1949.
7 M. BENSE, *Aesthetica IV,* op. cit., p. 165.

La vision de Norbert Wiener est elle-même une vision en rupture avec l'époque newtonienne. L'après-guerre est une période fertile pour la science qui se sort à peine de l'effort de guerre. Conscients des implications tragiques de la recherche dans la technologie de l'armement, Wiener et le groupe interdisciplinaire des conférences de Macy aux États-Unis – s'avancent sur de nouveaux terrains d'idées.

Dans « le hasard est une notion scientifique », la préface de *Cybernétique et société*[8], Wiener propage l'idée d'un renversement de vision sur le monde. Jusqu'alors, la physique newtonienne qui a dominé de la fin du XVII$^e$ jusqu'à la fin du XIX$^e$ siècle, décrit un univers dans lequel chaque chose advient à partir de lois concises, et dans lequel le futur dépend strictement du passé. Mais pour Wiener, les expérimentations qui doivent permettre de tester ces lois sont imparfaites. Elles ne peuvent jamais vérifier un tel ensemble de lois «jusqu'à la dernière décimale »[9]. Ludwig Bolzmann en Allemagne et Josiah Willard Gibbs aux Etats-Unis, tous deux physiciens – proposent une nouvelle idée stimulante pour pallier les limites de la vision newtonienne.

Comme le précise Wiener, le physicien James Clerk Maxwell considérait déjà des mondes faits d'une très grande quantité de particules et qui devaient alors être traités de façon statistique. Quant à Bolzmann et Gibbs, ils prouvent que la statistique, comme science de la distribution, peut à la fois être employée pour des systèmes très larges, mais peut aussi être étendue à l'étude de systèmes très réduits. La différentiation de ces systèmes ne repose plus alors sur un ensemble de lois causales. Il s'agit, grâce à la statistique (des positions et des vitesses) de s'intéresser à la répartition des éléments qui composent ce système. Autrement dit, la physique doit alors considérer « l'incertitude et le caractère accidentel des événements. »[10]

Conséquence de la « révolution de la physique du XX$^e$ siècle » par Gibbs, le hasard devient un instrument scientifique prépondérant pour de décrire le monde par la probabilité. La physique épouse le principe d'incertitude – c'est ce que montre l'étude du mouvement brownien[11], une description du mouvement aléatoire d'une particule immergée dans un fluide et dont Max Bense reproduira une image dans *Aesthetica* (Fig. 1). Le hasard est alors pensé comme une partie fondamentale de la Nature.

---

[8] Norbert WIENER, *Cybernetics and Society : The Human Use of Human Beings*, London, Free Association Books, 1989 [éd. orig. Houghton Mifflin, 1950].
[9] Norbert WIENER, *Cybernétique et société : L'usage humain des êtres humains*, Paris, Points, 2014.
[10] « Cela signifie simplement que nous ne connaissons pas parfaitement les conditions initiales, mais que nous avons une certaine idée de leur répartition. En d'autres termes, la partie fonctionnelle de la physique ne peut éviter de considérer l'incertitude et le caractère accidentel des événements. Ce fut le mérite de Gibbs de mettre en œuvre, pour la première fois, une méthode scientifique précise pour prendre cet aspect contingent en considération. » *Ibid.*
[11] Mouvement brownien ou processus de Wiener : dont Norbert Wiener propose une définition mathématique en 1923.

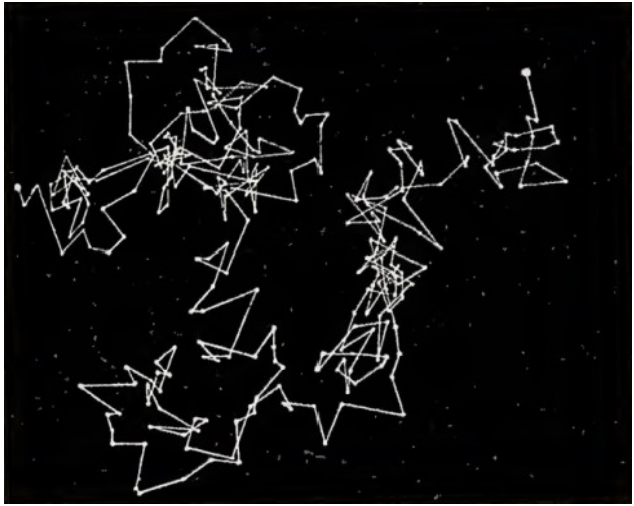

Fig. 1, « Mouvement moléculaire brownien. Un modèle structurel de base des processus thermodynamiques », reproduction in M. Bense, *Aesthetica : Einführung in die neue Aesthetik*, op. cit.

L'intérêt d'un tel changement de perspective, explique Wiener, c'est que l'on n'a plus affaire à un univers spécifique conçu comme un tout, mais « que les réponses aux questions que l'on se pose peuvent être trouvées dans un grand nombre d'univers similaires. »[12] C'est le principe d'un monde contingent. De plus, pour Gibbs, l'univers qui vieillit tend à l'entropie. Autrement dit l'univers et tous les systèmes qui le constituent tendent à se détériorer et à perdre leurs distinctions, passant de l'état le moins probable (une organisation dans laquelle des distinctions et des formes existent) à un état de chaos et de similitude. Dans cet univers « l'ordre est le moins probable, alors que le chaos est le plus probable. »[13]

À la cybernétique qui résout le problème de la communication et du contrôle – la tâche de développer une nouvelle science pour concevoir des systèmes dont l'évolution manifesterait une tendance à « l'accroissement de l'organisation. » Mais comment du point de vue des sciences Norbert Wiener en arrive-t-il à cette théorie de la communication, théorie essentielle de la cybernétique ? À quelle vision du monde, la cybernétique en particulier nous renvoie-t-elle ?

À l'instar de la physique statistique, la théorie de la communication est une théorie probabiliste. Conçue sur les bases du principe d'entropie de Gibbs, elle s'oppose à la tendance de la nature pour le désordre. Les messages émis entre émetteur et récepteur sont étudiés par rapport à un degré d'ordre ou de désordre mesuré en terme statistique. Wiener et Shannon s'intéressent au calcul de la quantité d'information contenue dans un message composé de lettres[14] et qui est transmis à travers un schéma de communication. Si une série de signes choisis au hasard ne transmet aucune information particulière, car la série est alors chaotique, le message « acquiert son sens du fait qu'il est sélectionné parmi un grand nombre de schémas possibles »[15]. Ce concept d'information repris par Bense dans *Aesthetica*, en particulier dans le livre II, est compris dans une théorie statistique de l'information.

---
[12] N. Wiener, *Cybernétique et société*, op. cit.
[13] *Ibid.*

Par extension, Max Bense pense les processus esthétiques comme des processus statistiques. Pour Bense qui reprendra directement la terminologie d'entropie pour discuter par exemple des œuvres de Max Bill (Fig. 2) ou de Henri Michaux[16] (Fig. 3), « la théorie de l'information et la théorie de la communication constituent *un pont entre physique et esthétique* ». Bense se propose d'employer des formules mathématiques pour interpréter des œuvres non figuratives. Tandis qu'il décrit la peinture concrète de Bill comme la répartition d'un contraste local au sein une structure régulière, il voit dans le tachisme des « mouvements » de Michaux la manifestation d'une entropie, c'est-à-dire une augmentation du degré de désordre.

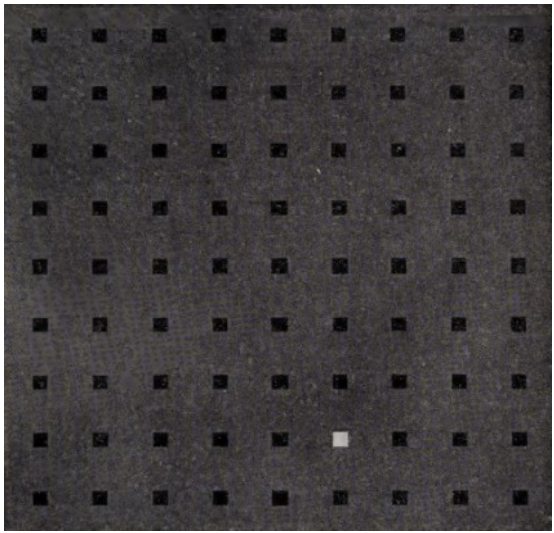

Fig. 2, Max Bill, *Weißes Quadrat*, 1946, huile sur toile, 70 x 70 cm, reproduction in M. Bense, *Aesthetica : Einführung in die neue Aesthetik*, op. cit.

---

14 M. BENSE, *Aesthetica : Introduction à la nouvelle esthétique,* op. cit, p. 287.
15 *Ibid.*, p. 296.
16 *Ibid.*, p. 225-227.

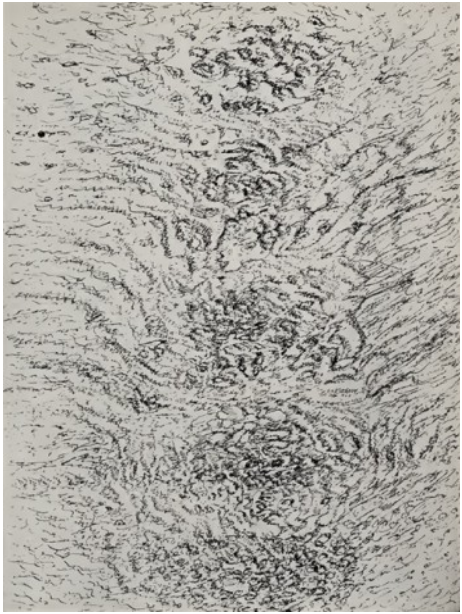

Fig. 3, Henri Michaux, *sans titre*, circa 1960, structure visuelle tirée de la série de dessins réalisée sous mescaline, reproduction in M. Bense, *Aesthetica : Einführung in die neue Aesthetik*, op. cit.

L'œuvre d'art se distingue comme organisation visuelle dans la mesure où – conçue en tant qu'une répartition d'information, elle peut être comprise comme le fruit d'une sélection inhabituelle, ou originale. En cela le processus esthétique est pensé comme un processus de différentiation d'une information entièrement désordonnée. Comme nous le verrons dans la partie suivante, et passant de l'art abstrait à la littérature, Bense poursuivra ces principes dans la programmation informatique afin d'établir des « méthodes numériques qui permettent de décrire un texte en recourant à des moyens qui sont pratiquement les mêmes que ceux dont se sert la thermodynamique pour décrire un gaz. »[17]

**Programmer ? Du langage scientifique au langage littéraire**

À travers sa construction d'une nouvelle théorie de l'art, ancrée dans les sciences naturelles et dans la cybernétique, le philosophe développe une esthétique qui s'éloigne de la matérialité de l'œuvre au profit de la cognition. Il se réfère notamment à l'introduction *à l'Esthétique* de Hegel, selon lequel « la pensée et la réflexion auraient pris le pas sur les beaux-arts ». L'explication de Wiener comme elle est citée par Bense : « *Information is information, not matter or energy* », illustre l'intérêt de ce dernier pour une esthétique qui ne se résume pas à un fait physique et qui ne se réduit pas non plus (exclusivement) à un fait de conscience. Bense cherche une nouvelle forme d'association entre civilisation et technique.

---

17 *Ibid.*, p. 352.

Le philosophe poursuit cette recherche dans l'atteinte à la séparation des sciences expérimentales et des sciences humaines. Si la technique prouve « la force génératrice de réalité des sciences expérimentales », celui-ci constate avec provocation que les sciences humaines, car elles analysent plus qu'elles ne génèrent – se limitent à « rester derrière les productions de notre intelligence ». Par un passage radical de l'analyse à la synthèse et avec l'aide de la machine informatique, Bense promet un revirement dans l'union de l'esthétique et de la technique. Cette union repose sur la nature informationnelle de l'œuvre.

En 1960, sa théorie d'une esthétique programmée n'est pas isolée. Bense est conscient des diverses expériences européennes dans la discipline de la théorie ou de l'esthétique de l'information (en musique avec Abraham Moles, également en journalisme au Centre International d'Enseignement Supérieur du Journalisme de Strasbourg, etc.). Reste à savoir à quel point celui-ci s'est rapproché des expériences réelles, menées alors dans les laboratoires de mathématiques de l'université de Stuttgart[18].

En 1958, Theo Lutz qui termine alors ses études en mathématiques dans l'institut universitaire de technologie, programme un ordinateur, le modèle Zuse Z22, dans le but de composer des poèmes stochastiques[19]. Tout en suivant une structure grammaticale pré établie, ces poèmes sont composés de phrases dans lesquelles les mots sont déterminés par un calcul d'aléatoire. Bense entend parler des tentatives de Lutz. Il lui propose de programmer l'ordinateur avec une sélection de mots de Franz Kafka. En 1959, Bense publie ensuite les *Textes stochastiques* de Lutz dans son magazine littéraire *Augenblick*[20]. La théorie de Bense rencontre un terrain épistémologique favorable au sein de l'institut de calcul de l'université[21].

Pour le philosophe, la littérature programmée trouve sa justification dans la théorie de la littérature. Bense élargit ce champ à la théorie des textes, englobant un éventail plus large dont les reportages, les longs métrages, les séries, les actualités, et la communication visuelle en général. Également inspiré par des auteurs tels que Gertrude Stein, James Joyce ou encore Francis Ponge, le philosophe reconnaît un certain nombre d'écrivains comme des précurseurs de la programmabilité du texte. Si Bense trouve en la synthèse de l'écrit la promesse d'une nouvelle discipline expérimentale, de quelle façon le philosophe établit-il une relation entre le langage scientifique et la littérature ?

---

18 Technische Hochschule Stuttgart.
19 Margit ROSEN, *et al.*, *Bense Und Die Künste: Eine Ausstellung Zum 100. Geburtstag Des Philosophen Max Bense (1910–1990)*, livret de l'exposition. Karlsruhe, ZKM, 2010.
20 Theo LUTZ, « Stochastische Texte », in Max BENSE (dir.), *Augenblick,* n°4*,* Baden-Baden, Agis Verlag, 1959. [consulté le 12 octobre 2020]. https://www.stuttgarter-schule.de/lutz_schule_en.htm
21 Christoph Hoffman, « Eine Maschine Und Ihr Betrieb: Zur Gründung Des Recheninstituts Der Technischen Hochschule Stuttgart (1956 – 1964) », in *Ästhetik Als Programm. Max Bense/Daten Und Streuungen*, Barbara Büscher (Hg.), Christoph Hoffmann (Hg.), Hans-Christian von Herrmann (Hg.). Berlin: Kaleidoskopien Schriftenreihe, 2004.

Dans *Aesthetica IV*[22], Bense fait également référence à la théorie de Rudolf Carnap. Carnap est un philosophe américain qui prône le positivisme logique. Sa philosophie repose sur la construction d'un langage pour la science, libéré de la métaphysique. En 1952, Carnap publie au MIT un « Aperçu d'une théorie de l'information sémantique »[23], un article dans lequel il présente une approche distincte de la théorie de la communication alors en vigueur avec Shannon et Weaver. Tandis que la théorie admise traite des quantités d'information comme une mesure de la rareté statistique d'un message, Carnap et Hiller construisent une théorie dans laquelle le concept d'information – réalisé par une phrase dans un système linguistique donné – est traité comme synonyme avec le contenu de cette phrase.

Bense élabore un exemple. En voici une restitution – prenons la conjonction suivante : « rouge et doux et atteignable », et considérons que les trois termes de cette conjonction peuvent être indépendamment modifiés en leur contraire (e. g. « bleu et doux et atteignable », ou « rouge et amer et atteignable », ou encore « bleu et amer et inatteignable », etc.). La question posée par Carnap et reprise par Bense est de savoir combien de propriétés cette proposition contient (le nombre de combinaisons possibles). Par principe combinatoire, le système engendre huit propriétés ou combinaisons distinctes[24]. Ce calcul entraîne une mesure de l'information sémantique contenue dans une proposition. Bense établit alors un parallèle entre la démarche scientifique de Carnap et la poésie concrète d'Eugen Gomringer. Pour Bense, la recherche et le formalisme de Carnap rejoignent l'idée de la *constellation* trouvée chez le poète.

Pour Gomringer, la *constellation* est une configuration simple qui a pour unité le mot. De la même façon qu'un dessin relie des étoiles pour former un signe, la constellation en poésie compose des groupes de mots qui forment un ensemble polysémique. À l'instar de Carnap, Gomringer ouvre la pensée du texte à une pensée combinatoire.

---

22 M. BENSE, *Aesthetica IV,* op. cit, p. 67.
23 Traduction de l'auteur. Rudolf CARNAP et Yehoshua BAR-HILLEL, « An Outline of a Theory of Semantic Information », *Research Laboratory of Electronics*, n° 247, Cambridge, MIT*,* 1952.
24 Si Q désigne le nombre de propriétés, alors $Q = 2^3 = 8$.

> La constellation est ordonnée par le poète. Il détermine l'aire de jeu, le champ ou la force et suggère ses possibilités. Le lecteur, le nouveau lecteur, saisit l'idée du jeu et se joint à lui [au poète].[25]

En écrivant à partir d'un processus de simplification formelle et avec un nombre réduit de formes minimales, Gomringer construit à la fois un « objet de pensée et un jeu d'idée »[26].

Le raisonnement logique, la brièveté des formes, la production du sens en fonction de la modélisation sémantique, retiennent de façon évidente l'attention de Bense. En termes de programmation, il y a bien pourtant une différence importante entre l'approche du poète et la structure d'un algorithme. Tandis que le sens du poème se dégage de la liberté donnée au lecteur de parcourir la surface de la page, le programme, pour engendrer les différentes combinaisons possible, doit suivre un ensemble d'instructions finies et précises. Pour programmer des combinaisons de termes dans une conjonction donnée, il serait nécessaire de décrire dans un langage de programmation chacune des opérations qui permettraient *in fine* de lister les combinaisons possibles. Il serait nécessaire également de décrire en premier lieu le répertoire des termes à être appelés dans la combinaison[27].

Si Bense crée un parallèle entre la modélisation de l'information sémantique chez Carnap et le jeu poétique, concret, chez Gomringer, la vision esthétique qu'il propose dans *Aesthetica IV* ne comprend pas encore la programmation comme la traduction d'opérations en processus informatique. Aussi l'emploi du terme « programmer » dans *Programmer le beau*, dont l'usage paraît singulier, témoigne d'un écart entre la vision du philosophe et ce que deviendra véritablement la recherche d'une esthétique algorithmique parmi les lecteurs ou étudiants de Max Bense.

On peut reconnaître cependant chez Bense une intuition qui nous oriente vers une vision des images programmées. Dans l'avant dernier chapitre titré « Textes visuels », celui-ci propose une définition qui ouvrirait à un passage entre une théorie textuelle générale à une théorie générale de l'image[28]. Par l'approche statistique et informationnelle, et par une réflexion sur « la plus petite unité d'image indivisible » le philosophe esquisse le principe désormais bien connu du *pixel*. L'effort entrepris avec *Aesthetica IV*, de traiter de façon formelle de l'information, du texte et de la sémantique – offre un terrain remarquable pour se projeter dans une théorie esthétique de l'information appliquée à l'image.

---

25 Traduction de l'auteur. Eugen GOMRINGER, « From Line to Constellation », in Mary Ellen Solt (dir.), *Concrete Poetry: A World View*, Bloomington, Indiana University Press, 1968 [éd. orig. « Vom vers zur concrete poesie », in Max BENSE (dir.), *Augenblick*, n° 2, Baden-Baden, Agis Verlag, 1954].
26 Traduction de l'auteur. Eugen GOMRINGER, *The Book of Hours and Constellation*, New York, Villefranche-sur-mer, Frankfurt-am-Main, Something Else Press, Inc., 1968.
27 Gaëtan ROBILLARD et Alain LIORET, « A Vision without a Sight: From Max Bense's Theory to the Dialectic of Programmed Images », in Celestino Soddu, Enrica Colabella (dir.), *XXII Generative Art – GA2019*, Roma, Domus Argenia, 2019, p. 138-149.
28 M. BENSE, *Aesthetica : Introduction à la nouvelle esthétique*, op. cit, p. 411.

**Nouvelles images, pour une esthétique générative**

L'étude du fond Elisabeth Walther-Bense au Zentrum Für Kunst und Medientechnologie de Karlsruhe, et plus particulièrement la partie de l'archive dédiée à la revue *Rot* [Rouge]*,* révèle une activité intense du philosophe dans l'édition. Le rapport entre Max Bense et le monde des images semble plus complexe que ce qui ressort de l'analyse de *Programmer le beau*[29], qui d'une certaine façon passe à côté de la théorie de l'image. Si la poésie concrète, que Bense crée ou publie, est une forme littéraire, c'est aussi certainement l'art de distribuer des signes dans les deux dimensions de la page. L'observation des pratiques d'édition à travers la revue *Rot* révèle la relation que Bense entretient avec les images et avec les artistes.

En mai 2019, dans le cadre d'une bourse de recherche de courte durée, j'ai eu l'opportunité de passer un mois entre le Laboratoire Compart de l'Université de Brême et le ZKM[30]. Ayant convenu d'un rendez-vous, je me rends au centre d'archive du ZKM afin de compléter mes recherches sur le Computer Art, accompagné d'un certain nombre d'interrogations sur l'activité intellectuelle de Max Bense. J'y découvre le fond d'Elisabeth Walther-Bense, femme et coéditrice du philosophe. Le fond se constitue d'une correspondance complète des écrits de Whalter avec artistes et théoriciens. Le fond comprend un journal personnel ainsi que tous les manuscrits concernant ses articles et publications (dont un segment important sur la sémiotique).

Outre son travail avec Max Bense sur les revues *Semiosis* et *Augenblick,* le fond contient également toutes les parutions et les documents (maquettes, matériel de reproduction, correspondance) de l'*Edition Rot*, revue expérimentale tenue par le couple. Le fond comprend d'autre part le travail de photographie d'Elisabeth Walther, des bobines de film qu'elle a tournées, notamment à l'occasion de voyages, documents dans lesquels se mêlent le professionnel et le personnel. De plus le ZKM dispose de la bibliothèque de recherche des deux éditeurs et philosophes.

---

29 M. BENSE, *Aesthetica IV,* op. cit.
30 Bourse de recherche de courte durée 2019, Deutscher Akademischer Austauschdienst, Office allemand d'échanges universitaires. Encadrant : Frieder Nake, Laboratoire Compart, Université de Brême. [Consulté le 13 octobre 2020]. https://compart.uni-bremen.de

Le matériel éditorial de l'*Edition Rot* est un matériel abondant en documents. Lors de ma venue en 2019, dans la salle de travail attenante à la salle principale de l'archive, plusieurs empilements de boîtes devaient encore être triés (Fig. 4). C'est dans cette salle que j'ai pu consulter et saisir des informations sur les cinquante-sept numéros de *Rot*.

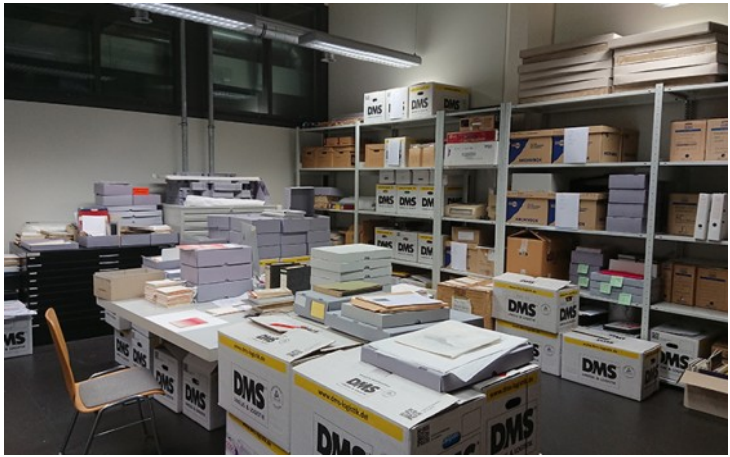

Fig. 4, Gaëtan Robillard, vue de la salle de travail de l'archive du ZKM, 2019.

Parallèlement à la publication d'*Aesthetica IV* en 1960, la revue *Rot* est lancée à Stuttgart. Elle réunira progressivement poésie concrète, sémiologie, typographie et cybernétique. Parmi le matériel à trier, un *flyer* datant de 1966 présente la revue ainsi :

> La série Rot est destinée à la publication de littérature et de travaux graphiques expérimentaux. Le terme d'expérimentation n'est pas défini de façon restrictive. Nous incluons tout ce qui peut être créé à partir de l'hypothèse d'un concept esthétique théoriquement accessible, y compris le contenu de la forme et le but de la production artistique. La littérature et les travaux graphiques expérimentaux se rapportent à toutes les techniques topologiques abstraites et concrètes, stochastiques et aléatoires. De plus, [la série] ne se limite pas aux méthodes de production naturelles d'un individu créatif mais compte également avec les productions artificielles de systèmes informatiques électroniques.[31]

Avec son format carré et sa couverture rouge conçue par Hansjörg Mayer (Fig. 5), *Rot* est reconnaissable au premier regard. Marquée par un usage de la typographie Futura sans majuscules, la revue compte cinquante-sept cahiers publiés de 1960 à 1997. L'édition incarne les activités du Cercle d'intellectuels de Stuttgart qui s'est formé autour de Max Bense et Elisabeth Walther. Parmi les auteurs de la revue, on peut trouver (sans ordre spécifique) : Jean Genet, Abraham A. Moles, Charles S. Pierce, Hansjörg Mayer, Georg Nees, Yona Friedman, Francis Ponge, Dieter Roth, Haroldo de Campos, et bien d'autres encore. La typologie de publication est riche tant pour les textes que pour les images : poésie en vers, prose, poésie concrète, philosophie, sémiologie, essai, théâtre, photographie, peinture, dessin, typographie, bande dessinée, diagramme, dessin ou texte généré par ordinateur...

---

31 Traduction de l'auteur. Max BENSE et Elisabeth WALTHER, feuillet publicitaire, 1966. ZKM | Center for Art and Media Karlsruhe / Elisabeth Walther-Bense Estate / ZKM-01-0129-02-0988.

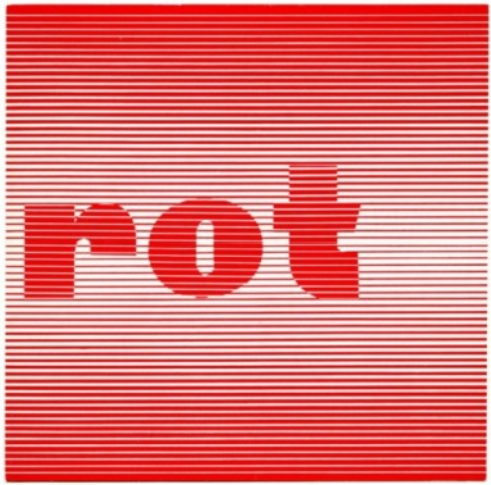

Fig. 5, Hansjörg Mayer, *Rot*, 1965, design de la couverture, ZKM | Center for Art and Media Karlsruhe / Elisabeth Walther-Bense Estate / ZKM-01-0129-02-0991-a

En proportion, la revue se centre principalement sur la poésie concrète et alterne fréquemment avec des numéros dédiés à la sémiologie et aux arts visuels. Trois cahiers, le n° 6, le n° 19 et le n° 50, sont explicitement orientés vers la programmation des formes esthétiques, avec respectivement des textes générés par Max Bense (1961), des dessins génératifs de Georg Nees accompagnés d'un texte de Bense prenant l'allure d'un manifeste (1965), et la poésie visuelle et programmée de l'américaine Carole Spearin McCauley (1972). Il faut également noter que le n° 8 présente un texte d'Abraham A. Moles intitulé « Premier manifeste de l'art permutationnel », accompagnée de la reproduction d'une image de Vasarely[32] (1962).

D'une manière générale, les cahiers *Rot* présentent un réseau intellectuel lié par un intérêt pour le texte, le signe typographique et la sémiologie. Compte tenu de l'implication de Max Bense dans la genèse du Computer Art, il peut paraître surprenant que le principe d'une esthétique programmée, avec trois numéros spécifiques, n'apparaisse que de façon mineure. Mais en feuilletant les cahiers un à un, il apparait clairement que l'expérimentation *Rot* est fertile. Sans jamais se répéter, la relation texte-image progresse de façon dynamique et il n'est pas rare qu'un numéro entier soit d'ailleurs consacré à un seul artiste ou à une série visuelle[33]. Si Bense a écrit de nombreux textes sur les beaux-arts[34], il est aussi un éditeur d'images. De plus Elisabeth Walther était sémioticienne et également photographe. Elle a pu prendre un rôle important sur la reproduction de photographies dans les cahiers *Rot*[35]. A-t-elle sensibilisé Bense au domaine de la photographie ? Une étude plus approfondie de l'archive permettrait certainement de faire le jour, au moins en partie, sur le rôle d'Elisabeth Walther dans la proximité de Bense avec le monde des images.

---

32 Abraham MOLES, *Erstes manifest der pemutationellen kunst*, Rot n° 8, Stuttgart, Max Bense, Elisabeth Walther, 1962.
33 Par exemple, dans l'esprit du Nouveau Réalisme avec le travail en photographie et collage de Reinhold KOEHLER, *Schtrottstempel*, Rot n° 27, op. cit., 1966.
34 M. ROSEN *et al.*, *Bense Und Die Künste*, op.cit., 2010

Ce qui transparait à travers l'étude du matériel éditorial, c'est une acceptation du monde visuel dans toutes ses nuances, allant de l'abstraction du signe typographique jusqu'à l'empreinte photographique du monde réel. Mais quel a été le rapport ou la posture du Bense-éditeur avec le matériel visuel produit *artificiellement* avec l'aide du système informatique ? Les archives du ZKM révèlent les maquettes de divers numéros dont le numéro n° 19, rendu célèbre pour son inauguration de l'*esthétique générative* ; une première pour la publication des images programmées dans le monde artistique.

Dans le n° 19 publié en février 1965 sous le titre *Computer-Grafik*[36], *Rot* inaugure l'esthétique générative. La première partie du numéro se consacre à un texte et à un ensemble de dessins informatiques de Georg Nees, ingénieur-mathématicien et artiste travaillant alors chez Siemens à Erlangen en Allemagne. D'un point de vue historique, ces dessins comptent parmi les premières images générééś par des algorithmes programmés sur un ordinateur[37]. En parallèle de cette publication, Bense organise un évènement public à la *Studiengalerie* de l'université de Stuttgart. Le philosophe y présente les dessins de Nees. Frieder Nake, également mathématicien et artiste pionnier du Computer Art, présent ce jour-là, rapporte l'évènement comme la première exposition d'art-informatique, « deux mois avant la célèbre exposition Howard Wise à New York »[38].

Dans les deux premières pages de la revue, Georg Nees présente un texte « sur les programmes d'infographie stochastique ». Chaque graphique se compose à partir de paramètres aléatoires. La répétition de paramètres aléatoires « produit l'improbabilité esthétique des graphiques ». Après un court paragraphe, Nees présente cinq parties écrites sous la forme d'un pseudo-code. Chaque partie concerne une ou deux images que l'on découvrira dans les pages suivante : « 8-coin: (image 1) », « 23-coin: (image 2) »[39], etc., jusqu'à l'image n° 6. En texte clair – sans spécification d'aucun langage de programmation – les descriptions présentent des instructions succinctes et programmables qui génèrent des résultats graphiques.

---
35 Par exemple le n° 60 dans lequel elle signe la photographie en 1995 ; Max Bense est alors décédé.
36 Georg NEES et Max BENSE, *Computer-grafik*, Rot n° 19, op. cit., 1965.
37 « Computer-Grafik (Nees 1965) », in Frieder NAKE (dir.), *Compart. Center of excellence digital art*. [consulté le 17 novembre 2019]. http://dada.compart-bremen.de/item/exhibition/325
38 L'exposition à la *Studiengalerie* et la publication de Rot n ° 19 datent de février 1965.
39 Traduction de l'auteur. G. NEES et M. BENSE, *Computer-grafik*, Rot n° 19, op. cit., 1965.

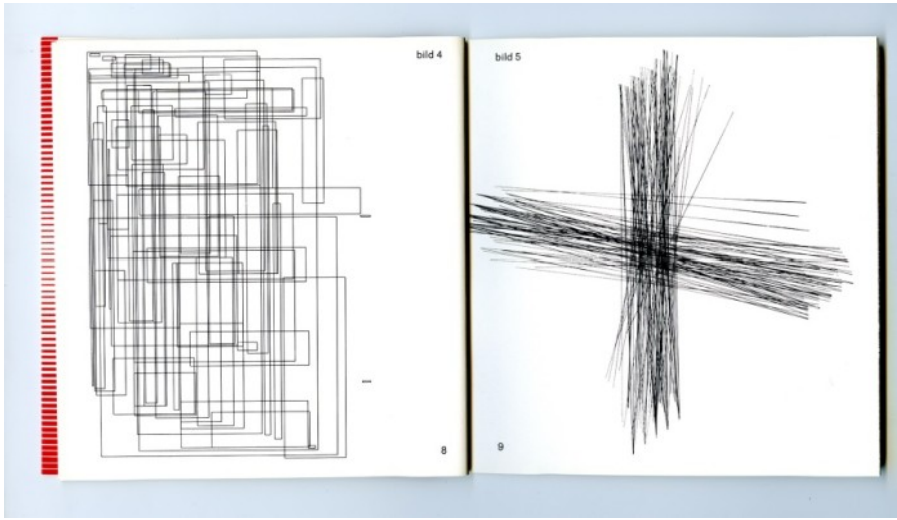

Fig. 6, Georg Nees, *Computer-Grafik*, in Rot n° 19, 1965, ZKM | Center for Art and Media Karlsruhe / Elisabeth Walther-Bense Estate / ZKM-01-0001-W-1014-f

Pour les images qui suivent, la mise en page est systématique : le titre de l'image (par exemple « image 1 »), l'image elle-même, en pleine page et de haut en bas, et le numéro de page du cahier. Les images présentées sont toutes faites de segments tracés à l'encre noire avec un trait d'une épaisseur d'environ 0,5 mm. Ce sont des lignes verticales, horizontales ou obliques. Les formes sont organisées sous forme de motifs dans une grille apparemment répétitive, ou bien en pleine page, à partir d'une variation de densité de lignes formant une composition abstraite (Fig. 6). Si une grande partie de l'œuvre de Nees a été réalisée à l'aide d'un *pen plotter*, la revue ne mentionne pas la technique de réalisation.

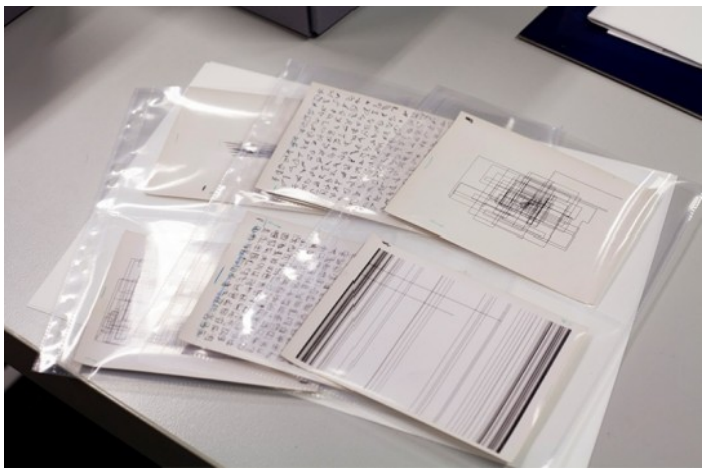

Fig. 7, Georg Nees, six images ou graphiques de la série in *Computer-Grafik*, Rot n° 19, matériel éditorial. ZKM | Center for Art and Media Karlsruhe / Elisabeth Walther-Bense Estate / ZKM-01-0129-02

En regardant de plus près les archives du fond Walther-Bense, je découvre des éléments précieux concernant la façon dont le cahier *Rot* n° 19 a été édité. Des reproductions des six images qui y sont publiées peuvent être trouvées sur papier glacé (Fig. 7), comportant diverses annotations au crayon bleu et au stylo vert de la main de Bense. Pour la reproduction de *23-Coin* (Fig. 8) qui présente une grille de signes géométriques aléatoires, les dernières formes en bas de l'image sont raturées. La signature de Nees qui se trouve en bas à gauche est également raturée. Deux lignes bleues et une flèche verte verticale située dans la marge de gauche semblent indiquer une instruction pour rogner l'image dans la disposition finale du cahier.

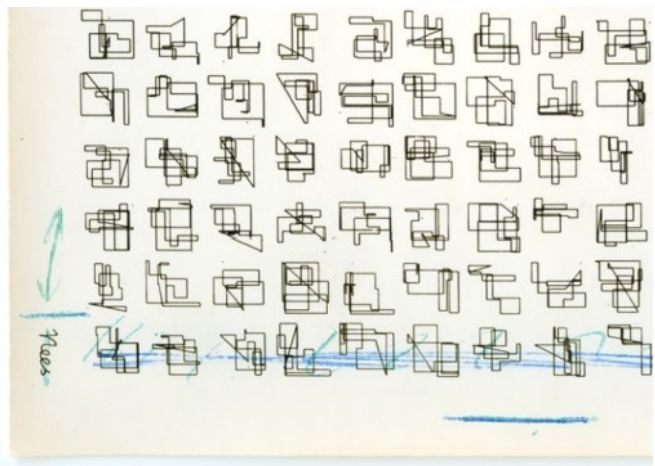

Fig. 8, Georg Nees, *23-Ecke*, 1965, détail d'une reproduction raturée et annotée par Max Bense, in Rot n° 19, 1965, ZKM | Center for Art and Media Karlsruhe / Elisabeth Walther-Bense Estate / ZKM-01-0129-02-0009-r

Quelles étaient les intentions de Bense ? Est-ce parce que le nom de Nees figurait en signature que la dernière ligne de figures géométriques a été supprimée de l'édition ? Ou était-ce simplement pour rapprocher la grille de signes du format carré de *Rot*, omettant la signature par la même occasion ? Bien que la réponse exacte reste inconnue, ce geste simple de recadrage de l'image témoigne de la relation de Bense à l'auteur et à la catégorie nouvelle des images programmées. Dans ce cas, c'est comme si les signes géométriques abstraits générés par l'algorithme de Nees fonctionnaient comme des caractères qui devaient être sélectionnés de nouveau suite à leur réalisation par le calcul.

D'une part la machine qui exécute le programme sélectionne des valeurs obtenues par le calcul de nombres pseudo-aléatoires, et d'autre part Bense qui exerce sa vision d'éditeur sélectionne de nouveaux les éléments de l'image à paraître. Ce geste de raturage rappelle de manière significative l'intérêt de Bense pour la réalisation de l'information comme

« expression d'une fonction de sélection »[40]. De plus, ceci révèle qu'à travers la parution inédite de ces images dans le n° 19, le but de Bense n'était peut-être pas de présenter Nees au public en tant qu'artiste, mais plutôt de démontrer l'application expérimentale de ses théories dans le domaine purement visuel.

La seconde partie du n° 19 fait paraître les « Projets d'esthétique générative » de Max Bense. Souvent présenté comme un manifeste des débuts du Computer Art, ce court texte de trois pages entend clarifier une méthode expérimentale de production d'états esthétiques par génération computationnelle[41].

Bense s'appuie sur la grammaire générative de Chomsky et se focalise sur la réalisation « artificielle » d'une structure esthétique programmable. Si à travers ses parutions, Bense avance sur le terrain nouveau des images programmées, l'étude de la revue montre que sur ce terrain, l'activité du philosophe est indissociable d'une interaction au sein d'un groupe d'intellectuels et d'artistes : le Cercle de Stuttgart. La théorie esthétique qu'il construit progressivement influence de jeunes chercheurs qui au sein des laboratoires universitaires ou industriels, expérimentent et programment les premières images du Computer Art.

**Frieder Nake : une dialectique entre algorithme et image**

Pour Max Bense, le langage mathématique permet la transition des processus physiques du monde naturel aux principes productifs du monde technique et artificiel. L'esthétique générative a elle pour rôle de formuler ces principes tout en portant l'attention sur la production d'états esthétiques par le biais de programmes. Par cette même occasion, *la nouvelle esthétique* reconfigure le rôle de l'auteur en médiateur des conditions auxquelles la machine doit obéir pour matérialiser un dessin ou une image.

Dans la publication de Georg Nees décrite plus haut, la juxtaposition de la description des algorithmes avec les images pose question. Etant donné cette proximité entre l'écrit et le visuel, comment analyser et interpréter l'image ? S'agit-il de retrouver la logique formelle qui se cache derrière celle-ci? Quels sont les outils critiques qui permettraient une évaluation de l'algorithme en simultané avec l'évaluation du résultat esthétique ? La thèse présentée ici est qu'une image programmée à l'aide d'un algorithme, qu'elle soit scientifique ou artistique, constitue une unité dialectique. C'est précisément parce que l'algorithme et l'image sont indissociables qu'ils constituent ensemble une catégorie spécifique d'image. L'image existe à la fois comme algorithme et comme matériau visuel.

---

[40] Traduction de l'auteur. M. BENSE, *Aesthetica IV,* op. cit.
[41] « Par esthétique générative, on entend la récapitulation de toutes les opérations, des règles, théorèmes, dont l'application à toute une série d'éléments matériels faisant office de signes rend possible par l'intermédiaire de ces derniers la production délibérée et méthodique d'états esthétique (qu'il s'agisse de répartitions ou de configurations) ». M. BENSE, *Aesthetica : Introduction à la nouvelle esthétique,* op. cit, p. 453.

La tension dialectique entre algorithme et esthétique n'est pas dans formulation de Bense ; on la trouve plutôt dans la théorie et l'enseignement de Frieder Nake[42]. Etudiant en mathématiques au début des années soixante, il assiste alors aux conférences du philosophe et s'introduit dans le Cercle de Stuttgart. En novembre 1965, neuf mois après la publication et la présentation des expériences graphiques programmées à la *Studiengalerie* de Bense, Nake et Nees exposent ensemble leurs travaux à la galerie du libraire Wendelin Niedlich (Niedlichs Buchladen und Galerie), en utilisant le titre *Computer Grafik* (très légèrement différent du titre du cahier *Rot* n° 19). Pour Frieder Nake, cette exposition marque un début de carrière artistique qui le mènera ensuite à une reconnaissance internationale à partir de 1968[43].

Aujourd'hui professeur à l'Université de Brême, Frieder Nake poursuit une réflexion sur la naissance d'images d'une nouvelle nature, en particulier dans le contexte technologique, scientifique et philosophique des années soixante. Dans une publication récente[44], il oppose la production visuelle-iconique qu'il relie à la pensée artistique ou picturale avec la production numérique-symbolique liée au contexte d'ingénierie dans lequel sont produits les premiers dessins par ordinateur. Il propose en particulier de voir chez Nees une transaction singulière entre ces deux régimes d'image, pour aboutir à un art d'essence algorithmique. Nake, pour qui « le Computer Art est un art conceptuel »[45], met en évidence la tension dialectique entre l'algorithme et l'image engendrée par le calcul :

> Lorsque l'ordinateur exécute la description, il lit cette description d'une façon particulière qui est la sienne : il réalise exactement ce que la description lui demande de faire, et rien d'autre. Lire, c'est toujours interpréter. L'ordinateur, lors de la lecture du texte opérationnel, en fait une interprétation. Absolument différente de notre interprétation, l'interprétation de l'ordinateur est une détermination :

---

42 Frieder NAKE, « Introduction to Digital Media », *Compart. Center of Excellence Digital Art*, [Consulté le 17 novembre 2019]. https://compart.uni-bremen.de/teaching/wintersemester-2019-20/introduction-to-digital-media
43 avec les expositions *Cybernetic Serendipity* à Londres, et *Tendencies 4: Computers and Visual Research* dans le cadre de la biennale *New Tendencies* à Zagreb. En 1970, le travail de Nake sera présenté dans l'exposition collective de la 35ᵉ Biennale de Venize : *Ricerca e Progettazione : Proposte per una esposizione sperimentale*.
44 Frieder NAKE, « Georg Nees & Harold Cohen: Re:tracing the origins of digital media », *Digital Art through the Looking Glass: New strategies for archiving, collecting and preserving in digital humanities*, Krems a.d. Donau, Edition Donau-Universität, 2019.
45 Traduction de l'auteur. Frieder NAKE, « Paragraphs on Computer Art, Past and Present », *CAT 2010 London Conference,* 2010, p. 57.

aucune liberté n'est permise. L'ordinateur interprète en déterminant la seule et unique interprétation qui a un sens algorithmique.[46]

La programmation joue un rôle de description formelle qui précède la matérialisation de l'image. Et à l'opposé du calcul – à l'opposé de la description algorithmique et de sa lecture par la machine – à la surface, l'image prend vie. En devenant visible, l'œuvre se manifeste dans le monde matériel et, parce qu'elle est visible, elle devient l'objet d'une expérience sensible. Dans cette perspective, l'étude du travail artistique de Nake offre un terrain d'analyse.

L'œuvre de Frieder Nake repose sur la programmation de dessins géométriques ou de grilles colorées composants des ensembles abstraits. Les travaux qu'il mène dans les années soixante présentent des espaces à la fois construit et enchevêtrés. L'*Hommage à Paul Klee* (Fig. 9) témoigne de l'expression d'une fidélité de Nake pour l'art de Klee. Nake s'inspire de l'œuvre *Chemin principal et chemins latéraux* (1929). Cependant tout en faisant hommage à la peinture, Nake diffère. Il se concentre moins sur la subdivision régulière et arithmétique d'une surface que sur la décomposition d'un espace potentiellement infini ou ouvert. Le programme agit sur deux niveaux : un premier processus dessine une grille à la forme aléatoire, puis un processus secondaire engendre des états visuels dans chaque cellule de la grille, produisant le dessin d'un maillage à l'équilibre entre ordre et désordre.

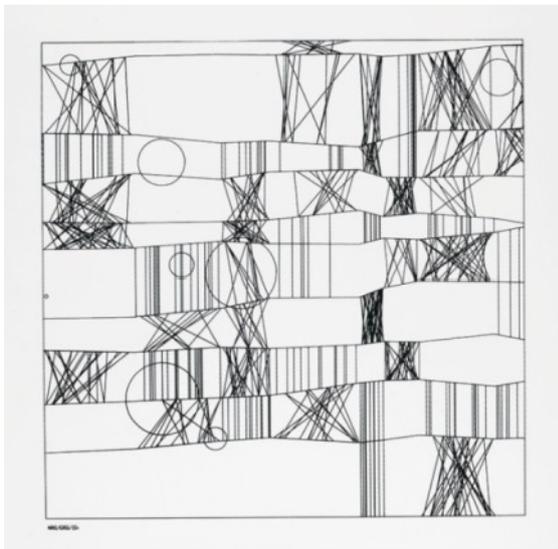

Fig. 9, Frieder Nake, *Homage à Paul Klee*, 1965, sérigraphie (orig. *pen plotter* sur papier), 49,2 x 49,2 cm, Brême, Kunsthalle, collection Herbert W. Franke.

---

46 *Ibid.*

Le caractère de l'essai mobilise l'image dans une dimension expérimentale. Bien que programmée, la dynamique du dessin est bifurquée au sein même du programme par une combinaison entre variable aléatoire et définition de lois de distribution statistique. Chaque exécution du programme résulte potentiellement en un nouvel équilibre. Car elle est programmée, l'œuvre est non finie. De plus le programme est toujours paramétrable et reprogrammable *ad infinitum*. Sa structure même peut être amenée à changer de niveau. Par certaines opérations de recadrages ou de choix d'encre pour la réalisation du dessin, l'artiste peut encore intervenir sur l'image une fois le calcul achevé. L'œuvre se réalise donc à l'intersection du calcul et de la matérialisation de l'image.

Dans ses textes récents, Nake théorise une œuvre-processus[47], déterminée essentiellement par la dynamique du changement. Le régime dialectique de l'image algorithmique étudiée dans la théorie et l'œuvre de Frieder Nake paraît anticiper la nature complexe des images du XXIe siècle, en particulier dans le domaine de la visualisation et de l'image de synthèse. Pour rendre le monde visible, l'image doit être actualisée par le calcul. Cette nature technique, et au fond statistique et informationnelle de l'image, pose question. Ceci par exemple déplace notre compréhension des images sur un autre terrain que celui de la reproductibilité pensée par Walter Benjamin. L'apparition de ce problème pourra d'ailleurs être développée dans une nouvelle étape de recherche. Ici, l'articulation esthétique entre le visible et le calculable nous entraîne à une réflexion sur l'imagerie scientifique d'aujourd'hui, en lien le tournant environnemental.

**Réflexions sur l'image en art et en science**

En 2019, grâce à la résidence AIRLab[48] et à une invitation de l'Institut Universitaire Européen de la Mer[49], j'ai eu l'opportunité de développer un travail de recherche et de création portant sur le rapport entre mathématique et environnement, et en particulier sur le calcul de vagues en mouvement. Intitulé *La vague dans la matrice* (Fig. 10), ce cycle de travail a débuté par une phase d'immersion et d'enquête au sein de laboratoires ou d'instituts qui s'efforcent de modéliser des phénomènes naturels et d'en représenter les résultats numériques. À cette occasion, j'ai été amené à rencontrer des chercheurs de l'institut France Energies Marines.

---

47 F. NAKE, « Georg Nees & Harold Cohen: Re:tracing the origins of digital media », op. cit.
48 Résidence AIRLab, Comue Lille Nord de France.
49 Invitation du festival RESSAC, Université Bretagne Ouest, dans le cadre des quatre-vingts ans du CNRS.

Cette association d'intérêt privé et public, basée à Brest, se concentre sur l'analyse environnementale et le développement de solutions pour les nouvelles énergies.

Jean-François Filipot est le directeur scientifique et chercheur pour la modélisation des vagues dans l'institut. Sa recherche se concentre sur la caractérisation des sites et le diagnostic des forces des vagues de tempête contre les infrastructures techniques. Il étudie par exemple des phénomènes climatologiques exceptionnels liés au réchauffement climatique. Pour cette raison, les images produites à travers la recherche de l'institut engagent à une réflexion sur la façon dont les sciences du climat engendrent des signes dans le monde contemporain.

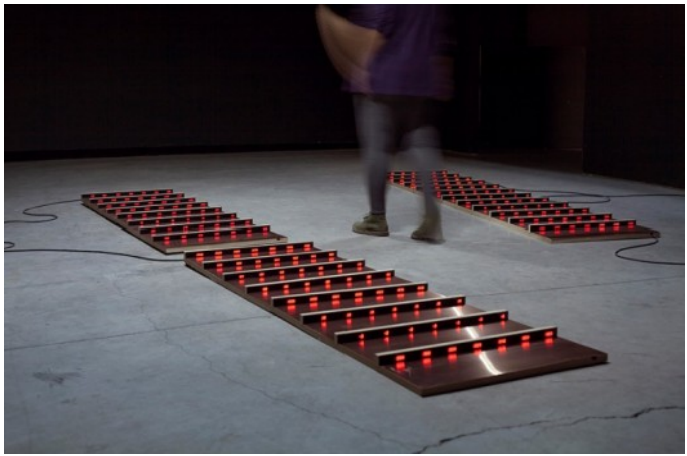

Fig. 10, Gaëtan Robillard, *La vague dans la matrice*, installation comprenant 33 nano processeurs, afficheurs 7 segments, cuivre, bois, programme de calcul scientifique modifié, dimensions variables, 2019.

À travers l'enquête, l'équipe de France Energies Marines a partagé avec moi un programme de visualisation utilisé pour étudier la qualité des états de mer en des lieux et à des instants précis. Chacun des ensembles de données lus par le programme correspond à une bouée particulière et à une période de temps allant de un à plusieurs mois. La méthodologie employée est basée sur une approche spectrale. Les spectres mesurent la distribution de l'énergie des vagues en fonction de leurs fréquences et de leurs directions. Les données enregistrées peuvent alors être représentées par des cartes de température (Fig. 11). Les données ne sont pas un enregistrement direct du mouvement dans sa totalité, mais plutôt un enregistrement de ses constituants à un instant donné. L'image est une image statistique.

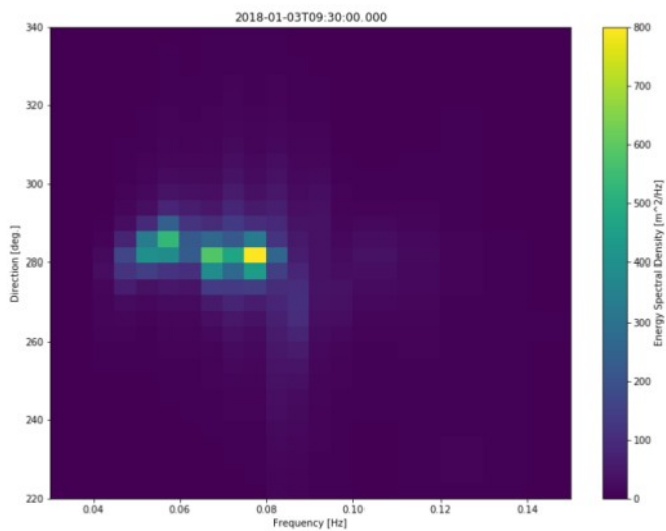

Fig. 11, Visualisation des spectres sur la bouée DATAWELL (France Energies Marines), Phare de la Jument, France, 3 janvier 2018. Rendu Gaëtan Robillard, 2020.

Au Laboratoire d'Océanographie Physique et Spatiale où l'enquête a été étendue, les états de la mer sont également modélisés par approche statistique. Fabrice Ardhuin, l'un des membres du LOPS, parle de la vague comme d'une signature dont on ne considère pas l'unicité mais la variance[50], ou son espace de *variation* pour revenir à un terme artistique. Ceci pose tout à fait les enjeux de la visibilité de ce phénomène et de sa considération en tant que population ou agent interagissant avec le climat.

Dans une seconde étape de l'enquête, je me suis concentré sur un nouveau programme conçu par l'équipe de France Energies Marines destiné à reconstruire de façon dynamique, par approche générative, les champs de vagues relatifs aux données statistiques (Fig. 12). L'algorithme charge les données et calcule une valeur de départ, la phase, qui est une valeur aléatoire et qui a pour effet de distribuer les vagues de façon chaotique. Chaque fois que le programme est exécuté, une nouvelle valeur de phase est définie – un nouveau champ de vague est construit. *In fine*, le programme génère une grille de hauteurs, résultante d'une opération sur la somme des multiples ondes. Cette grille peut être ensuite transformée en une image en trois dimensions. La librairie Matplolib, une librairie propre au langage de programmation Python est utilisée afin de *rendre* ou de *tracer* (*to plot*) les images.

---

50 Entretien du 4 juin 2019 avec Fabrice Ardhuin au LOPS, IUEM, Brest.

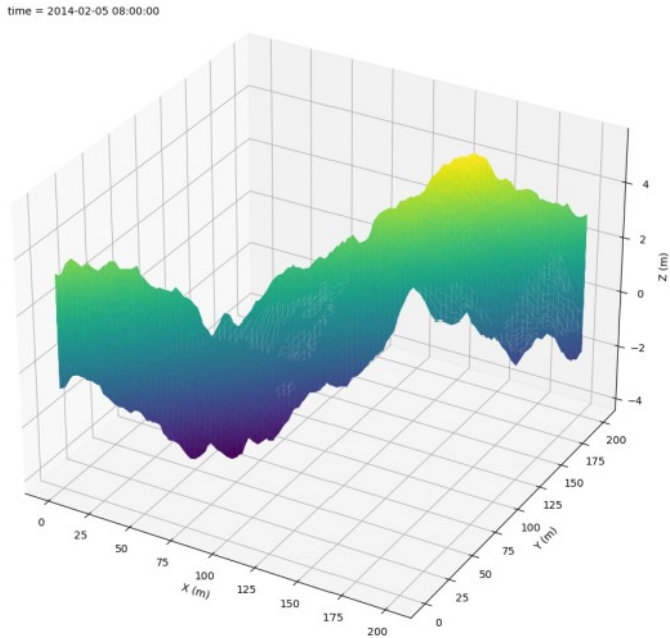

Fig. 12, Image 3D sortie du programme wavegen.py conçu par France Energies Marines, données : Bouée CANDHIS, Les Pierres Noires, France, février 2014. Le programme original permet de rendre une image bidimensionnelle, le rendu 3D présenté ici a été élaboré spécialement pour les besoins de l'enquête. Rendu Gaëtan Robillard, 2020.

Le fait qu'une valeur aléatoire soit calculée à chaque nouvelle exécution du programme marque le caractère probabiliste et génératif du champ de vagues résultant. Bien que le programme fonctionne sur une base de données déjà enregistrées, les vagues qu'il génère à chaque démarrage sont toujours différentes (étant donné que la date reste inchangée). Si la reconstruction est utilisée pour produire des connaissances sur les vagues et le climat, le *plot* – l'image rendue, reste le témoin d'une réalité probable, contingente ou indéterminée. L'image présente un champ de vague qui en réalité n'a pu ni être mesuré dans toute sa durée, ni observé par l'œil humain. Indissociable de son caractère algorithmique, la simulation est considérée ici comme constituante d'une image du monde, une composante à la fois historiographique et virtuelle qui participe à la compréhension de notre environnement en mutation. Modifié, le programme a été ensuite utilisé pour engendrer un calcul de valeurs numériques dans l'installation *La vague dans la matrice.*

Les images de Frieder Nake et l'exemple de visualisation scientifique présenté ici ont en commun le hasard comme élément constituant. Tandis que Nake distribue des signes abstraits à partir de lois de probabilités, la visualisation scientifique distribue des ondes afin de produire une surface dynamique et réaliste. L'image programmée issue du Computer Art et celle issue de la climatologie sont des images-processus. Elles comportent en elles une part d'indéterminé. La généalogie commune à ces deux registres iconographiques remonte aux bouleversements que connaît la science durant le XXe, et la théorie esthétique de Max Bense permet de recouper ces deux registres. Ceci ouvre à un champ d'étude visuelle qui nécessite de combiner à une analyse visuelle une autre analyse de type algorithmique.

**Conclusion**

Max Bense est-il un visionnaire ? A-t-il anticipé le régime algorithmique des images ? L'esprit expérimental, littéraire et graphique que l'on trouve chez Max Bense, son attachement inconditionnel à Norbert Wiener, sa volonté de rompre avec les conventions – rattachent le philosophe à un sens du regard en rupture avec les conventions artistiques de son temps. Dans *Programmer le beau*, Bense occupe une position avant-gardiste dans le domaine littéraire. Il propose aussi et dès 1960 un premier pas vers une théorie générale, statistique et informationnelle de l'image. Sa vision fait écho à des recherches pionnières conduites par des étudiants ou des chercheurs au sein des laboratoires de Stuttgart et Erlangen, nouvellement équipés en machine de calcul. Bense se positionne au centre du Cercle de Stuttgart, motivé par les fondations d'une nouvelle esthétique liée à la technique informatique.

La théorie esthétique de l'information puis l'esthétique générative publiée dans *Rot* trouve également des applications pour décrire les images algorithmiques du Computer Art d'une part et l'imagerie scientifique contemporaine d'autre part. Cette théorie fournit des moyens d'analyse (statistique, hasard, générativité) pour deux iconographies de natures pourtant éloignées. Ceci conduit à penser l'étendue de la vision de Bense, comprise entre sciences expérimentales et sciences humaines.

Cependant, la théorie esthétique de l'information de Max Bense prend source dans une révolution scientifique annoncée avant lui par le physicien et cybernéticien Norbert Wiener. Du point de vue de l'image programmée, même si Bense voit dans l'analyse statistique de l'œuvre d'art la possibilité d'un nouveau modèle, ce n'est qu'ensuite, avec Frieder Nake, qu'apparaît de façon distincte un discours critique portant sur la dialectique entre calcul algorithmique et image sensible. Alors que l'esthétique générative semble avoir été abandonnée aussitôt par Max Bense, Frieder Nake poursuit une théorie de l'œuvre ouverte et de l'image-processus.

D'autre part, les limites de la vision d'une civilisation vouée à sa réalisation rationnelle et technique nous apparaissent aujourd'hui pleinement. D'ailleurs contemporaine de la pensée de Bense, la théorie critique d'Adorno et Horkheimer[51] oppose la réalisation humaine de l'humanité à la calculabilité du monde. Désormais, la dimension algorithmique des images se développe en parallèle d'une certaine logique de la productivité rationnelle et capitaliste de la société, ce qu'un artiste comme Christophe Bruno a par exemple su révéler à travers des opérations artistiques sur les algorithmes de Google[52]. À cette perspective critique s'ajoute désormais la crise climatologique qui de façon urgente appelle d'autres philosophies à se développer.

Il n'en reste pas moins que la lecture de Bense amène à une critique des images algorithmiques de notre présent et des manières d'opérer sur elles. Face aux incertitudes d'un monde de plus en plus complexe, la question n'est peut-être pas tant de savoir si Bense répond à la figure d'un visionnaire, éclairant un chemin déterminé par avance. Il s'agirait plutôt de se demander comment celui-ci, par un saut esthétique, entraîne à une bifurcation dans des façons de faire et de penser les images.

**Remerciements**

---

[51] Theodor W. ADORNO, et Max HORKHEIMER, *La Dialectique de La Raison. Fragments Philosophiques*, Paris, Gallimard, Paris, 1974 [éd. orig. *Dialektik der aufklärung. Philosophische fragmente*, New-York, Social Studies Association, 1944].
[52] Christophe BRUNO, « Réseaux, art et logique », in David ZERBIB (dir.), *In octavo. Des formats de l'art*, Annecy, ESAAA Editions ; Dijon, Les presses du réel, 2015.